\documentclass[12pt]{article}

\usepackage{graphics,epsfig}
\renewcommand{\bottomfraction}{0.7}

\newcommand{\ci}[1]{\cite{#1}}
\newcommand{\lab}[1]{\label{#1}}

\newcommand{\ba}{\begin{eqnarray}}
\newcommand{\ea}{\end{eqnarray}}
\newcommand{\beqs}{\begin{eqnarray}}
\newcommand{\eeqs}{\end{eqnarray}}

\newcommand{\tbl}{\tablename}

\begin{document}
\markboth{O.V. Selyugin}
{The effects of the small-$t$ properties of hadronic scattering amplitude \\
on the determination its real part}


%
\title{The effects of the small-$t$ properties of hadronic scattering amplitude \\
on the determination its real part
}

 \author{
O. V. Selyugin\footnote{selugin@theor.jinr.ru}  \\
\begin{small}
 \emph{BLTPh, Joint Institute for Nuclear Research, Dubna, Russia} 
  \end{small}
}
%
%
\date{ }
\maketitle

\abstract{
   Taking into account the different forms of the Coulomb-hadron interference phase and the possible spin-flip contribution the new analysis of the experimental data of the proton-antiproton  elastic scattering  at $3.8 < p_L <6.0 \ $GeV/c and small momentum transfer is carried out.
   It is shown that the size of the spin-flip amplitude can be determined
   from the form of the differential cross sections at small $t$,
   and the deviation of $\rho(s,t)$ obtained from
  the examined experimental data  of the $p\bar{p}$  scattering
  from the analysis \cite{Kroll}, based on the dispersion relations, is conserved in all examined assumptions.
  The analysis of the proton-proton elastic scattering at $9 < p_L < 70 \ $GeV/c also shows the impact of the
  examined effects on the form of the differential cross sections.
  }



%
\section{Introduction}

        Hadronic high-energy physics has progressed tremendously since the early days of the S-Matrix theory
  and of Regge poles. Initially, it was believed that the imaginary part of the (spin-non-flip) amplitude was what really mattered in the forward direction because of the constraints coming from unitarity \cite{Preazzi02}.
  Eventually, this led to the birth and growth of the pomeron philosophy. As time went by and the analysis got more and more refined and higher and higher energies were explored, people realized that the actual picture was considerably more complex. Analyticity soon showed that one could not do without a real part \cite{mart,roy} while polarization data proved that it was not possible to ignore spin complications. The long believed axiom that "spin complications disappear at asymptotic energies" has (so far) never found confirmation; alternatively, the asymptotic domain has never been reached (yet). In fact, it was shown that the spin-flip amplitude gives a non negligible contribution and indeed it was the polarization data that led to the first decline of the simple minded Regge pole picture.


   The measure of the $s$-dependence of the total cross sections $\sigma_{tot}(s)$
   and of $\rho(s,t)$ - the ratio of the real to imaginary part of the
 elastic scattering amplitude  is very important as they are connected
  to each other through   the integral dispersion relations
  \ba
  && \rho_{ \frac{pp}{p\bar{p}} }(E)  \sigma_{ \frac{pp}{p\bar{p}} }(E) =   \frac{A_{\frac{pp}{p\bar{p}}}}{p}
     + \frac{E}{\pi p}  \int_{m}^{\infty} dE^{'}p^{'} [\frac{ \sigma_{ \frac{pp}{p\bar{p}} }(E') }{E'(E'-E)} -
     \frac{ \sigma_{ \frac{p\bar{p}}{pp} }(E') }{E'(E'+E)}].
   \ea

       The validity of this relation can be checked at LHC energies \cite{Rev-LHC}. The deviation can point
      out to the  existence of a fundamental length at TeV energies \cite{Khuri1,Khuri2}.
      But for such a conclusion we should know with high accuracy the lower energy data as well.
   The difficulty
  is that we do not know the energy dependence of these amplitudes and
  individual contributions of the asymptotic non-dying spin-flip
  amplitudes. As was noted in \cite{wak}, the spin-dependent part
  of the interaction in $pp$ scattering is more important than
  expected and a good
  fit to the data in the Regge model requires an enormous number of poles.

 As we  do not know exactly,  from a theoretical
  viewpoint,  the dependence  of the
  scattering amplitude on $s$ and $t$ one, it is usually  assumed
   that the imaginary and real parts of the spin-nonflip
  amplitude behave  exponentially  with the same slope.
   Similarly, one assumes the
  imaginary and real parts of  the spin-flip amplitudes (without the
  kinematic factor $\sqrt{|t|}$) to have an analogous $t$-dependence in the
  examined domain of momenta transfer.
  Moreover, one assumes  energy independence of
  the ratio of  spin-flip  to  spin-non-flip parts   at small $t$.
    All this is  our theoretical uncertainty.

\label{sec:figures}
\begin{figure}
\begin{center}
\includegraphics[width=0.45\textwidth] {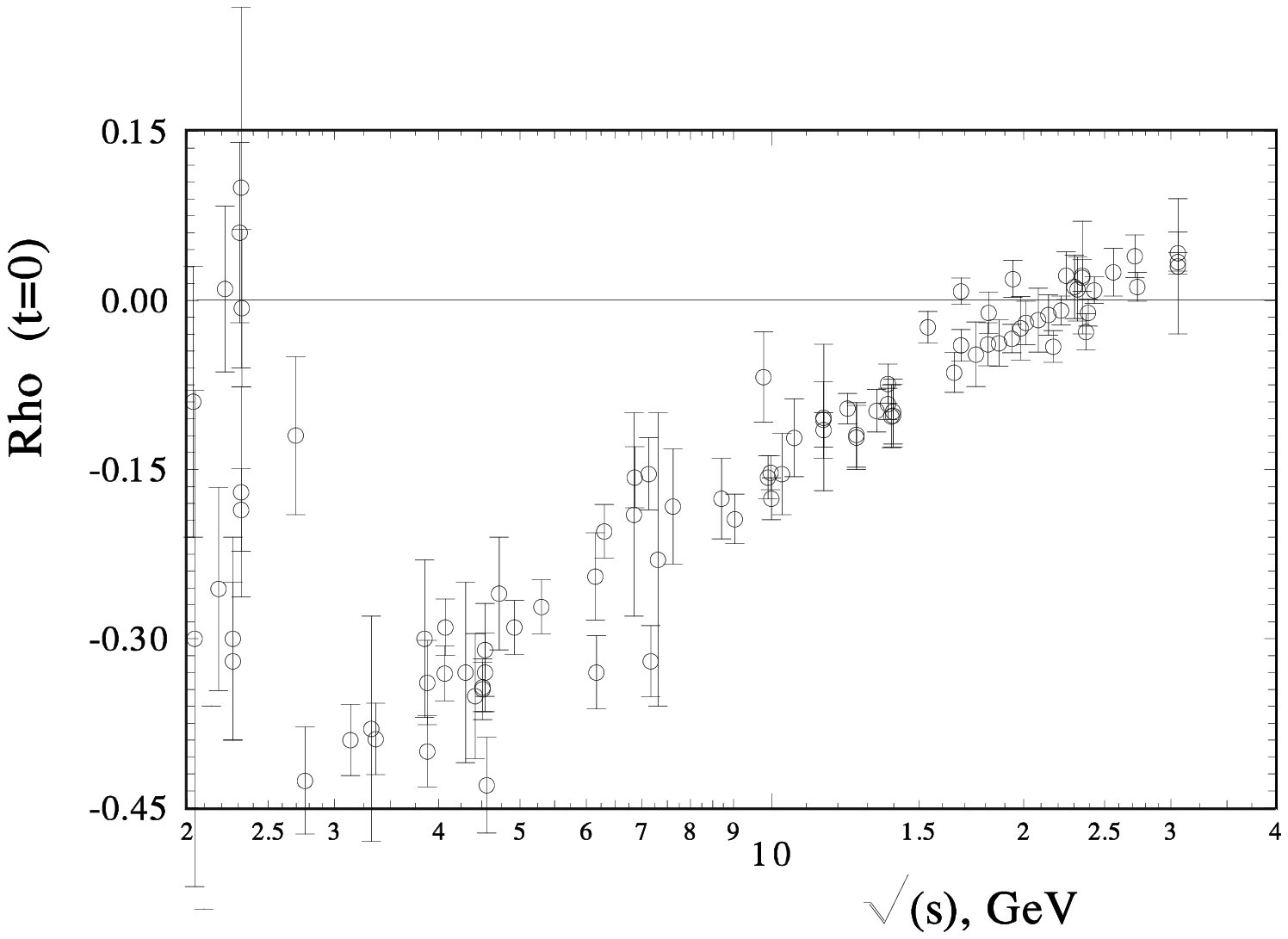}
\includegraphics[width=0.45\textwidth] {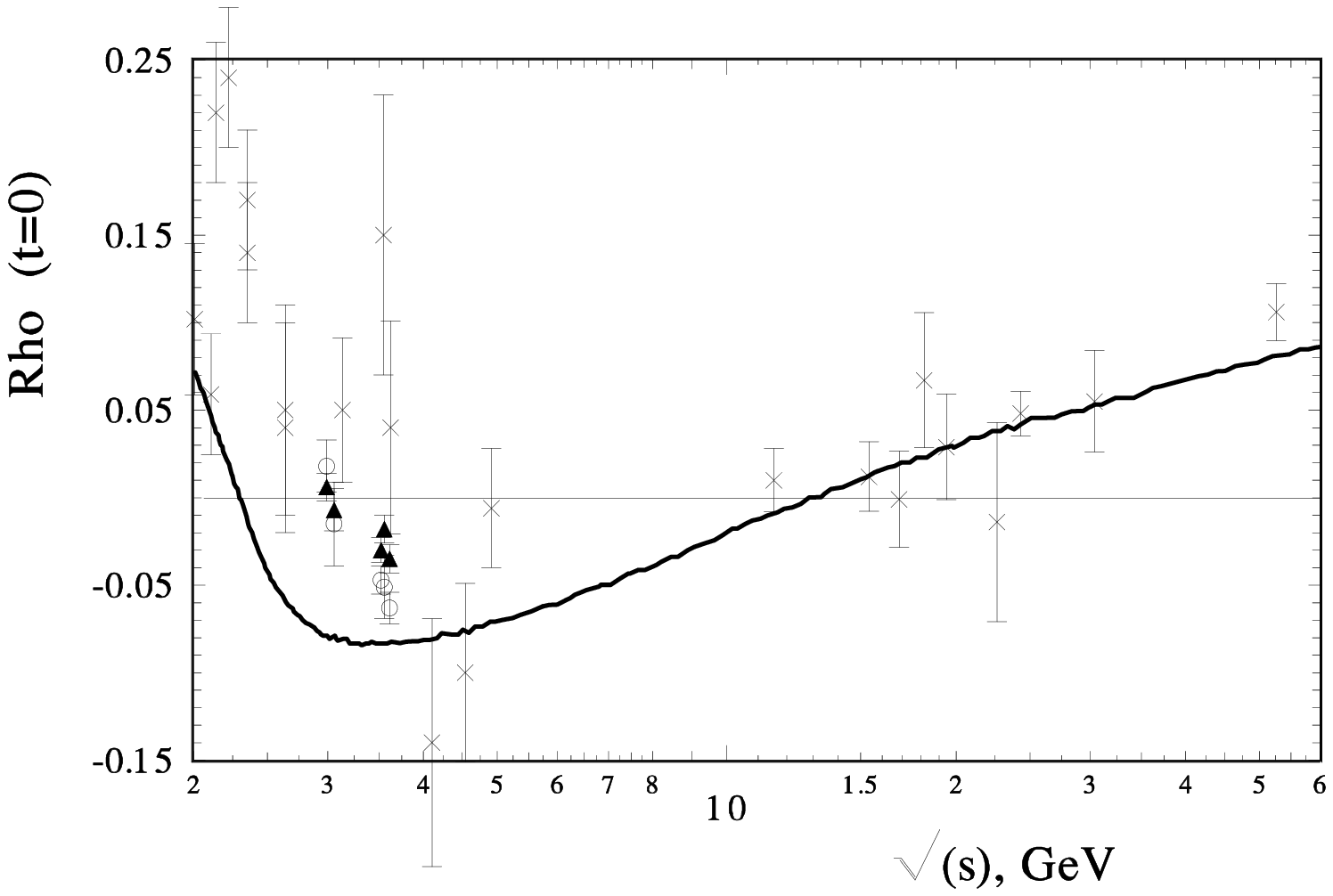}
\end{center}
 \caption{
 $\rho(s,t=0)$ - the ratio of the real to imaginary part of the
 elastic scattering amplitude of a(top plot) for  $pp$ scattering
  [experimental points - the world database);
  with curves of the different models
and b(bottom plot) for
  $p{\bar p}$ scattering at low
 energies (triangles - the fit with fixed $\sigma_{tot}$ and circles -
 the fit with "free $\sigma_{tot}$"  
 the curve is the dispersion relation description for the
  $p \bar{p}$ scattering 
 } \label{Fig_2}
\end{figure}

   In this paper we consider in the greatest detail the situation concerning $\rho(s,t)$.
   The model we propose takes into account all known features of the near forward
   proton proton and proton antiproton data, slopes of the spin-non-flip
   and of spin-flip amplitudes, total cross sections, ratios of the real to the imaginary
   forward amplitudes and Coulomb-nuclear interference phase
   where the form factors of the nucleons are also considered.
  Our main difficulty is  due to  the
 lack of  experimental data at high energies and small  momenta transfer with high accuracies.

  The non-trivial procedure of the extraction of the size of $\rho(s,t)$ from the experimental data on the
  differential cross sections shows the semi-phenomenological properties of $\rho(s,t)$.
  Its size is dependent on  some theoretical assumption \cite{SCP-EPJ08}.
  For example, a significant discrepancy  in  the experimental measurement of
 $\rho $ was found by the UA4 and UA4/2 collaborations at $\sqrt{s}=541 $~GeV.
But a more careful extrapolation \cite{SelyuginPL94} to $t=0$   shows
that there is no real contradiction between these measurements and
gives for this energy $\rho(\sqrt{s}=541 {\rm GeV}, t=0) = 0.163 $,
  the same as in the previous phenomenological analysis  \cite{SelyuginYF92}.

        Of course, we have plenty of  experimental
  data in the  domain of small  $t$ at low energies
   $3  < p_L < 100 \ (GeV/c)$.  Unfortunately,  most of these  data
   come with large errors.
 The extracted sizes of $\rho(s,t=0)$ contradict each other in the different experiments (Fig.1a) and give a bad $\chi^2$
  in the different models trying to describe the $s$-dependence of  $\rho(s,t=0)$
  (see, for example, the results of the COMPETE Collaboration
  \cite{COMPETE1a,COMPETE1b,COMPETE2}).
  It is of first importance that
   a more careful analysis of these experimental data gives in some cases an essentially different
  extrapolation for $\rho(s,t=0)$.
  For example, the analysis of the experimental data made in \cite{SelyuginYF92},
  which takes into account the
  uncertainty of the total cross sections ($3$-parameters fit) and the uncertainty of the Luminosity
  ($4$-parameters fit) gave a $\rho(s,t=0)$, which differs from the original values
  obtained by the experimental group,  by $25 \% $ on average.  
   For example, for $p_L = 19.23 \ $GeV/c
  the experimental work gave $\rho(s,t=0)=-0.25 \pm 0.03$ and for   $p_L = 38.01 \ $GeV/c
    $\rho(s,t=0)=-0.17 \pm 0.03$. The analysis with free $4$-parameters
    gave for these values:  $\rho(s,t=0)=-0.32 \pm 0.08$
   and $\rho(s,t=0)=-0.12 \pm 0.03$, respectively.
  This kind of picture was confirmed by
   the independent analysis of the experimental data \cite{Kuznetzov1,Kuznetzov2}
     $52  < p_L < 400 \ $(GeV/c)
    of Fajardo \cite{Fajardo}
   and Selyugin \cite{SelyuginYF92}. Both new analyses coincide with each other but differ from  the  original
    experimental determination.

          The data on proton-antiproton elastic scattering at
          $3.7  < p_L < 6.2 \ (GeV/c)$ are  most interesting. These experimental data have  high
          accuracies and give the extracted value of $\rho(s)$ with high
          precision  \cite{PL-Tro}. On Fig.1(b) these data are shown together with other experimental data
          and the predictions of the dispersion analysis carried out by Kroll \cite{Kroll}.
    The data for $\rho(s)$ essentially differ from the theoretical analysis.
    Hence, a more careful analysis of the original
    experimental data is required to take into account the different assumptions and corrections
    to the scattering amplitude.

 In this paper we will examine how different effects can influence the determination of $\rho(s,t)$.
 First we will consider the effect of the Coulomb-hadron interference phase (sections 2 and 3).
 Then we will consider spin effects (sections 4 and 5) and then, briefly, the effect of oscillations (section 6).

\section{The Coulomb phase and the hadron  form-factor}

 The electromagnetic amplitude can be calculated in the framework of QED.
   In the one-photon approximation
  and the high energy approximation, it can be  obtain \cite{bgl}
  for the spin-non-flip amplitudes:
  \begin{eqnarray}
  F^{em}_1 = \alpha f_{1}^{2} \frac{s-2 m^2}{t}, \  \ \
   F^{em}_3 = F^{em}_1;
  \end{eqnarray}

   and the spin-flip amplitudes:
    \begin{eqnarray}
   F^{em}_{4} = - F^{em}_{2}, \ \ \
  F^{em}_5 =  \alpha \frac{s }{2m \sqrt{|t|}} f_{1} f_{2}.
  \end{eqnarray}
  where $f_{1} (t)$ and $f_{2} (t)$ are the  Sachs form factors of the proton
  \begin{eqnarray}
  f_{1}(t) = \frac{4 m^2 -t (1+k)}{4 m^2 -t} G_{d}, \ \ \
  f_{2}(t) = \frac{4 m^2 k}{4 m^2 -t} G_{d},
  \end{eqnarray}
  with
   $ G_{d} = (1- t/\Lambda^2)^{-2} $;
  where $t=-q^2 $ GeV$^2$ is the momentum transfer, $\Lambda^2=0.71 $ GeV$^2$,
   $m$ is the proton mass, and $k= 1.793$ is the anomalous magnetic moment of the proton.

  The differential cross
  sections of nucleon-nucleon elastic scattering  can be written as the sum of different
  helicity  amplitudes:
\begin{eqnarray}
  \frac{d\sigma}{dt} =
 \frac{2 \pi}{s^{2}} (|\phi_{1}|^{2} +|\phi_{2}|^{2} +|\phi_{3}|^{2}
  +|\phi_{4}|^{2}
  +4 | \phi_{5}|^{2} ). \label{dsdt}
\end{eqnarray}

   Including  electromagnetic and hadronic interactions, every amplitude $\phi_{i}(s,t)$,
  can be expressed as
\begin{eqnarray}
  \phi_{i}(s,t) =
  F^{em}_{i} \exp{(i \alpha \varphi (s,t))} + F^{h}_{i}(s,t) ,
\end{eqnarray}
where
\begin{eqnarray}
  \varphi(s,t) =  \varphi_{C}(t) - \varphi_{Ch}(s,t).
\end{eqnarray}
   $ \varphi_{C}(t) $ will be calculated in the second Born approximation
 in order  to allow the evaluation   of   the Coulomb-hadron interference term $\varphi_{Ch}(s,t)$.
   The  quantity $\varphi(s,t)$
 has been calculated and discussed  by many authors.
 For  high energies, the first results were obtained
 by  Akhiezer,  Pomeranchuk \cite{akhi}
 for the diffraction on a black nucleus.
 Using the WKB approach in potential theory,  Bethe  \cite{bethe}
 derived  $\varphi(s,t)$   for proton-nucleus scattering
 \ba
  \varphi(s,t) = 2 \ln{(1.06/q a)},
\ea
 where
 the parameter, $a$, characterizing the range of the strong-interaction
 forces is defined by the size of a nucleus.
  After some  improvement  \cite{solov,rix}, the most
  important result was obtained by   Locher \cite{loch} and then by
  West and Yennie  using
  the Feynman diagram technique \cite{wy}.
  If  the hadron amplitude has the usual Gaussian form
  \begin{eqnarray}
   F^{h}(s,t) = h_{nf}(s) \ \exp{(-B(s) \ q^{2}/2)}, \end{eqnarray}
  where $h_{nf}$ denotes the "non-flip" amplitude,
   one gets the standard phase which is used in most experimental works,
\begin{eqnarray}
 \varphi(s,t) = \mp [\ln{(-B(s) t/2)} + \gamma],  \label{wyph}
\end{eqnarray}
where
 $B(s)/2$ is the slope of the nuclear amplitude,
 $\gamma$ is the Euler constant,
 and the upper (lower) sign
 corresponds to the scattering of particles with the same (opposite)
 charges.

   The influence of the electromagnetic form factor
  of  the scattered particles on
 $\varphi_{Ch}$
  in the framework of the eikonal approach was examined by  Cahn \cite{can}.
  He derived for $t \rightarrow 0 $ the eikonal analogue of \cite{wy} 
  and   obtained the phase, with take into account the form factor in the monopole form,
\begin{eqnarray}
\varphi_{Ch} (s,t)&=&\mp [\gamma +\ln{ (B(s)|t| /2)}
         + \ln{ (1 + 8/(B(s)\Lambda ^2))} \nonumber\\
    & & + (4|t|/\Lambda ^2)\ \ln{ (4|t|/\Lambda^2)} + 2|t|/\Lambda^2],
                                                 \label{Chane-ph}
\end{eqnarray}
where
$\Lambda =0.71 $ GeV$^2$. For the dypol form of the form factor
 it is need take some approximation with  $\Lambda =0.71/4 $ GeV$^2$.

The calculations of the phase factor beyond the limit $t \rightarrow 0$
  was carried out in \cite{selmp1}.

   In the absence of nuclear forces, we have
 only the Coulomb amplitude.
   In the second Born approximation there appear the additional phase,
   $\varphi(t)_{C}$
which are sufficiently complicated but
eventually exactly calculable (see refs. \cite{selmp1,selmp2,PRD-Sum}  for details).
 As a result, for the total Coulomb scattering amplitude, we have the eikonal
approximation of the second order in $\alpha (=e^2/hc)$
\begin{eqnarray}
F_C(q)&& =  F_C^{1B} + F_C^{2B}  =   -\frac{\alpha}{q^2}
         [\frac{\Lambda^4}{(\Lambda^2+q^2)^2}]
  [1+i\alpha(\{ \ln(\frac{\lambda^2}{q^2})  \ + \ \nu_s \}],
   \label{2B-C}
\end{eqnarray}
where
\begin{eqnarray}
   \nu_s =   A \ln(\frac{(\Lambda^2+q^2)^2}{\Lambda^2 q^2})
  + \ B  \ln(\frac{4 \Lambda^2}{(\sqrt{(4 \Lambda^2 +q^2}+q)^2}) \  + \ C. \label{e23}
\end{eqnarray}
 The coefficients $A,B,C$ are defined in \cite{PRD-Sum}.
\begin{eqnarray}
   A &=& \frac{q^2 (2\Lambda^2+q^2)}{\Lambda^4};    \ \ \
    C =  \frac{ 2\Lambda^4 - 17 \Lambda^2 q^2 - q^4}{(4\Lambda^2+q^2)^2}; \nonumber \\
   B &=& \frac{(\Lambda^2+q^2 )^2 [4\Lambda^4 (\Lambda^2 +7q^2 )+q^4 (10\Lambda^2+q^2 )]}{\Lambda^4 q ( 4\Lambda^2 +q^2 )^{5/2}}.
     \label{e23}
\end{eqnarray}

The impact of the spin of the scattered particles was analyzed in
\cite{bgl,lap} by using the eikonal approach for the scattering amplitude.
  Using the helicity formalism for high
  energy hadron scattering in \cite{bgl} it was shown that at
  small angles, all the helicity amplitudes have the same $\varphi(s,t)$.

\section{Impact of the Coulomb-hadron interference  phase}

\begin{table} 
\tablename{ 1\hspace{0.5cm} Proton-antiproton scattering (the phase dependence)}\\
{\begin{tabular}{@{}|c|c|c|c|c|c|c|c@{}}  \\  \hline
$p_{L}(\frac{GeV}{c})$ & N & $\rho_{exp.}$ &  $\rho_{\varphi(Born), n=1}$  &
$\rho_{\varphi(our), n=1}$ & $\rho_{\varphi(our), n-free }$  & n \\
\hline
3.702 & 134 & $+0.018 $ & $+0.046 \pm 0.005$  &  $+0.033 \pm 0.01 $ & $-0.004 \pm  0.005$ & 1.08\\
4.066 & 34 & $-0.015  $ &  $+0.038 \pm 0.008 $  &  $+0.026 \pm 0.01  $ & $-0.039 \pm 0.02 $ & 1.15\\
5.603 & 215&  $-0.047 $ &  $+0.004 \pm  0.003$  &  $-0.004 \pm  0.003 $ & $-0.037 \pm 0.015$ & 1.06\\
5.724 & 115&  $-0.051 $ &  $+0.014 \pm 0.01 $ &  $+0.002 \pm   0.004  $ & $-0.053\pm  0.012 $ & 1.12\\
5.941 & 140&  $-0.063 $ &  $-0.002 \pm 0.005$  &  $-0.012 \pm  0.004 $ & $-0.075 \pm 0.013$  & 1.13\\
6.234 & 34 &  $-0.06  $ & $-0.016 \pm 0.004 $  &  $-0.028\pm  0.02  $ & $-0.007 \pm 0.02 $ & 0.96 \\ \hline
\end{tabular}\label{ta1} }
\end{table}

 Let us  define the hadronic
spin-non-flip amplitudes as
\begin{eqnarray}
  F^{h}_{\rm nf}(s,t)
   &=& \left[F^h_{1}(s,t) + F^h_{3}(s,t)\right]/2; \label{non-flip}
    \end{eqnarray}
Equations (\ref{non-flip}) were applied at high energies and at small
momentum transfer, with the following usual assumptions
for helicity amplitudes:
 $\phi_{1}=\phi_{3}$, $\phi_{2}=\phi_{4} = 0 \ $;
 and, as usual, we neglect in this section the possible contribution of the spin-flip amplitude.

  We now compare  fits to
 the experimental  $p\bar{p}$-scattering \cite{PL-Tro,Disser,Disser-data,Land-Bron} data
  at low energies with different approximations
   for the Coulomb-hadron interference phase factor (see also \cite{Dubna-11}).
   Here we also include the additional normalization of the differential cross sections $n$
   which reflects the systematical errors of the luminosity.
    First, we use
   the simple  form of the phase, eq. (\ref{wyph}), with taking into account the $n=1$.
       The values obtained  for $\rho(s,t=0)$ are shown in the second column of Table 1.
   The results lie  above
the values  of $\rho(s)$ extracted    during the  experiments.


   If we take a slightly more complicated phase, eq.(\ref{Chane-ph}), with the dipole form factor     and our correction term,
   eq.\ref{2B-C},
   taking into account the two photon approximation and the dipole form factor,
   the new fitting procedure with $n=1$ gives the different sizes of $\rho(s)$
   (3 column  of Table 1). Now the results slightly decrease but still lie above
   $\rho_{exper.}$ for all examined energies.  (see Fig.1b).  If we take $n$ as a free parameter, the obtained size of $\rho$ more decrease as well,
     in two cases ($p_L = 3.7 $ and $ 4.066 \ $GeV/c) lie  slightly lower then the  $\rho_{exper.}$.
  However, in this case we obtain  large corrections of the normalization of the experimental
    data (up to $15\%$). It is shown that the form of the differential cross sections is
    far away from the simple exponential dependence.

\section{The slope of the hadron spin-flip amplitude}

  As it is not possible to calculate  exactly  the hadronic  amplitudes
  from  first principles,
  we have to resort to some assumptions for what concerns
  their form ($s$ and $t$  dependence) \cite{Sel-spin,PS-EPA02,CPS-EPA04}.
   Let us define the slope of the scattering amplitude as
   the derivative of the logarithm of the amplitudes with respect to $t$.
     For an exponential form of the amplitudes this  coincides
    with the usual slope of the differential cross sections divided by $2$.

 In most analyses, one makes the
  assumptions that the imaginary and real parts of the spin-non-flip
  amplitude have an exponential behavior with the same slope and that the
  imaginary and real parts of the spin-flip amplitudes, without the
  kinematic factor $\sqrt{|t|}$ \cite{sum-L},
   are proportional to the corresponding parts of the non-flip amplitude.

That is not so as regards the $t$ dependence
  was shown in Ref. \ci{soff}, where
  $F^{h}_{sf}$ is multiplied 
  by a phenomenological $t$-dependent function.
  Moreover, one usually  takes the ratio of the spin-flip parts to the spin-non-flip parts of the
  scattering amplitude  to be energy independent.
All this is our theoretical uncertainty \cite{CPS-EPA04,M-Pred}.
 In \cite{slope-MPL99} it was shown that in the case $B^{-}=2B^{+}$
  the contribution of the spin-flip amplitude
  can be felt in the differential
 cross sections of the elastic hadron scattering at small momentum transfer.

  According to the standard interpretation, the hadron spin-flip amplitude is
  connected with quark exchange between the scattering hadrons,
  and at large energy and small angles it can be neglected.
  Some models, which take into account  non-perturbative effects,
  lead to the non-dying hadron spin-flip amplitude \cite{mog2a,mog2b,mog2c,mog2d,CPS-PN04}.
  Another complicated question is related to the difference
  in phases of the spin-non-flip and spin-flip amplitude.

   We introduce the small-$t$
  spin non-flip  amplitude in the form
\ba
Im \ F^{h}_{nf}(s,t) \sim exp(B_{1}^{+} \ t), \ \ \
  Re \ F^{h}_{nf}(s,t) \sim  exp(B_{2}^{+} \ t),
\ea
and we define the "residual" spin flip amplitude where the kinematical vanishing at $t=0$
has been removed.
\ba
 Im  \tilde{F^{h}_{sf}}(s,t) =
 \frac{1}{\sqrt{|t|}} Im \phi^{h}_{5}(s,t) \sim  \ exp(B_{1}^{-} \ t);
 \ \ \
  Re \tilde{F^{h}_{sf}}(s,t) \  = \ \sim  \ exp(B_{2}^{-} \ t).
\ea
 At small $t$ ($\sim 0 \div 0.1 \ GeV^2$), practically all
 semiphenomenological analyses assume:
 $$ B_{1}^{+} \ \approx \ B_{2}^{+} \ \approx \  B_{1}^{-} \ \approx \ B_{2}^{-} . $$

    Actually, if we take the eikonal representation for the helicity
   amplitudes
\ba
F^{h}_{nf}(s,t) = - i p \int_{0}^{\infty} \ b \ d b  \ J_{0}(q b)
 [e^{\chi_{0}(s,b)} \  - \ 1 ],
\ea
\ba
F^{h}_{sf}(s,t) = - i p \int_{0}^{\infty} \ b \ db
 \ J_{1}(q b) \ \chi_{1}(s,b) \ e^{\chi_{0}(s,b)} .
\ea
where
\ba
\chi_{0}(s,b) = \frac{1}{2ip} \int_{-\infty}^{\infty} dz
 \   V_{0}(\vec{b}, z);
\ea
\ba
\chi_{1}(s,b) = \frac{b}{2ip} \int_{-\infty}^{\infty} dz
  \ V_{1}(\vec{b}, z)
\ea

  If the potentials $V_{0}$ and $V_{1}$ are
   assumed to have a Gaussian form
$$ V_{0,1}(b, z) \sim \ \int_{-\infty}^{\infty}\frac{1}{4B\sqrt{\pi B} }
   e^{\ r^2/(4B)} \ d z
 \ = \ \frac{1}{2 B} e^{- b^2/(4B)} $$
  where $r^2 = b^2+z^2$. In the first Born approximation
  $\phi^{h}_{1}$ and $\hat{\phi^{h}}_{5}$
   will have  the same slope
\ba
F^{h}_{nf}(s,t) &\sim&  \int_{0}^{\infty} \ b \ db
 \ J_{0}(q b) e^{- b^2/(4B)} /(2B)\  \nonumber  \ \ = \ \ e^{-B q^{2}}; \lab{f1a}
\ea
\ba
F^{h}_{sf}(s,t) &\sim& \int_{0}^{\infty} \ b^2 \ d b
 \ J_{1}(q b) \
   \ e^{-b^2/(4B) }/(2B) \ \ = \ 2 \ q \  B \ e^{-B q^{2} }. \lab{f5a}
\ea

  In this special case, therefore,
the slopes of
 the  spin-flip and  ``residual''spin-non-flip amplitudes are
  indeed the same.

 The first observation that the slopes don't coincide
 was made in \cite{predaz1,predaz2}.
  It was found from the analysis of the
 $\pi^{\pm} p \rightarrow \ \pi^{\pm}p $ and
  $pp \rightarrow \ pp $
reactions
  at $p_L \ = \ 20 \div 30 \ GeV/c $
  that the slope of the ``residual'' spin-flip amplitude is about
   twice as large as
  the slope of the spin-non flip amplitude. This conclusion was
  confirmed by the phenomenological analysis carried out in \cite{wak} for
  spin correlation parameters of the elastic proton-proton scattering
  at $p_L \ = \ 6 \ GeV/c$.

   The model-dependent analysis based on all the existing experimental
   data of the spin-correlation parameters above $p_L \ \geq
 \  6 \ GeV$
   allows  us to determine the structure of the hadron spin-flip
   amplitude at high energies and to predict its behavior at
   superhigh energies \cite{yaf-wak}. This analysis shows
   that the ratios
   $Re \ \phi^{h}_{5}(s,t) / (\sqrt{|t|} \ Re \ \phi^{h}_{1}(s,t))$ and
   $Im \ \phi^{h}_{5}(s,t)/(\sqrt{|t|} \ Im \ \phi^{h}_{1}(s,t))$
   depend on $\ s$ and $t$. At small momentum transfers,
   it was found that the slope of the ``residual'' spin-flip
    amplitudes is approximately
   twice the slope of the spin-non flip amplitude.

    Let us see what we obtain in the case of an exponential tail
  for the  potentials  \cite{PS-EPA02}.
 If we take
$$ \chi_{i}(s,b) \ \sim \ H \ e^{- a \ b}, $$
 and use the standard integral representation
\ba \int_{0}^{\infty} \ x^{\alpha-1} \ exp(-p \ x) J_{\nu}(cx)\ dx \
   = \ I_{\nu}^{\alpha}, \nonumber
\ea
with
\ba
 I_{\nu}^{\nu+2} \ = \ 2 p \ (2c)^{\nu} \ \Gamma (\nu + 3/2)
 1/[\sqrt{\pi} (p^{2}+c^{2})^{3/2}], \nonumber
\ea
 we obtain
\ba
 F_{nf} (s,t) = \int \rho \ d\rho \ e^{- a \ \rho }
  \ J_0 (\rho q) \ = \ \frac{a}{(a^2 + q^2)^{3/2}} \approx \
 \frac{1}{a\sqrt{a^2+q^2}} \ e^{-B q^2} \nonumber
\ea
with $B \ = \ 1/a^{2}$,
 where we have used the approximation
 $$ 1/(1+x) \ \sim  \ (1-x) \ \sim \ exp(-x). $$

For the ``residual'' spin-flip amplitude,
  on the other hand,
 we obtain
\ba \sqrt{|t|} \tilde{F_{sf}}(s,t) &=&
  \int b^{2} \ d b \ e^{- a \ b} \ J_1 (q b)  = \frac{3 \ a \ q}{(a^2 + q^2)^{5/2}} \
 \approx
 \ \frac{3 \ a  q \ B^{2}}{ \sqrt{a^2+q^2}} \ \ e^{-2 \ B q^2}
\ea
  In this case, therefore,
  the slope of the ``residual'' spin-flip amplitude exceeds the slope
 of the spin-non-flip amplitudes by a factor of two.

 Hence, a  long tail hadron potential implies a significant difference of the
  slopes of the ``residual'' spin-flip and of the spin-non-flip amplitudes.
   Note,  that the procedure
  of  eikonalization would lead to a further increase
  of the difference of these two slopes.

\section{Impact of the spin-flip contribution}

   In \cite{Runco}, it is claimed that the analysis of the low energy experimental data
   is not affected by  the spin-flip amplitude on the
   extracted value of $\rho(s,t)$. We believe  that this result
   must be checked up carefully and that  at low energies the size of the spin-flip
   amplitude is non negligible.
   In this work, we examine a simple model for the spin-flip
   amplitude and try to find its impact on the determination of $\rho(s,t)$
   from the low energy data of proton-proton scattering.
   Based on the analysis carried out in the previous section  we take
   the spin non-flip and spin flip amplitudes in the simplest exponential form
   \ba
       F^{h}_{nf}= h_{nf} \ [i+\rho(s,t=0)] \  e^{B^{+} t/2};
\ea
and
    \ba
       F^{h}_{sf}= \sqrt{-t}/m_{p} \ h_{sf} \ [i+\rho(s,t=0)] \  e^{B^{-} t/2},
\ea
with $B^{-}=2B^{+}$.
 and the full amplitudes will include the
   corresponding  electromagnetic parts and the Coulomb-hadron
   phase factors as analyzed previously.

  The results of our new fits of the proton-antiproton experimental data
  at $ p_L = 3.7 .. 6.2 \ $GeV/c are presented in Table 2.
  The changes of $\sum_{i}\chi_{i}^2$ after including the contribution
  of the spin-flip amplitude are reflected by the coefficient
\ba
 R_{\sum_{i}\chi_{i}^2} \ = \ \frac{\sum_{i}\chi^2_{i \ without  \ sf.}
 - K \sum_{i}\chi^2_{i \ with  \ sf.} }{\sum_{i}\chi^2_{i \ without \ sf.}}.
  \label{Rchi}
\ea
 where $K=(N-m_1)/(N-m_2)$ is the coefficient taking into account the difference
 of the number of free parameters ($m_1=3$ for the case without spin-flip contribution and
 $m_2=4$ with taking into account the spin-flip contribution).

\renewcommand{\bottomfraction}{0.7}

\begin{table}
\tbl{ 2  Proton-antiproton elastic scattering (the spin dependence)} \\
{\begin{tabular}{@{}|c|c|c|c|c|c|c@{}} \\  \hline 
 $p_{L(GeV/c)}$  &   N & $\rho_{exp.}$  &$R_{\chi^2}$
 &  $\rho_{model}$ & $h_{sf}$, GeV  \\
              &               &               &          &    &   \\ \hline
3.702 & 134 & $+0.018 \pm  0.014$ &  $7.3 \%$  &  $+0.055  \pm   0.001$ & $ 38.3 \pm 0.4$  \\
4.066 & 34 & $-0.015 \pm  0.024$ &  $22.5 \%$  &  $+0.061 \pm  0.02$ & $ 45.4 \pm 6.7$  \\
5.603 & 215&  $-0.047 \pm 0.017$ &  $3.1 \%$  &  $+0.002 \pm 0.005$ & $ 31.1 \pm 3.7 $ \\
5.724 & 115&  $-0.051 \pm 0.011$ &  $5.7 \%$  &  $+0.023 \pm 0.001$ & $ 30.8 \pm 0.4$  \\
5.941 & 140&  $-0.063 \pm 0.015$ &  $3.8 \%$  &  $+0.009 \pm 0.016$ & $ 38.6 \pm 0.4$  \\
 \hline
\end{tabular}\label{ta1} }
   \end{table}

\begin{table}
\tbl{ 3   Proton-proton elastic scattering (the spin dependence)} \\
{\begin{tabular}{@{}|c|c|c|c|c|c|c@{}} \\  \hline 
$p_{L(GeV/c)}$   & N & $\rho_{exp.}$  &$R_{\chi^2}$
 & $\rho_{model}$ & $h_{sf} $, GeV  \\
              &               &               &          &    & \\ \hline
9.4 & 34 & $-0.372 \pm 0.$ &  $ 1.2 \%$  &  $-0.413 \pm 0.02$ & $ 32.1 \pm 15.2$\\
18.9 & 67 & $-0.266 \pm  0.008$ &  $4.8 \%$  &  $-0.283 \pm 0.011$ & $ 7.8 \pm 1.7$ \\
38. & 65 & $-0.161 \pm   0.006$ &  $2.6 \%$  &  $-0.169 \pm 0.007$ & $ 6.9 \pm 1.7$\\
39.4 & 47&  $-0.195 \pm  0.015$ &  $0.02 \%$  &  $-0.201 \pm 0.01 $ & $ 15.76 \pm 5.$\\
40.6 & 65&  $-0.159 \pm  0.007$ &  $0.1 \%$  &  $-0.17 \pm  0.01  $ & $ 12.9 \pm 1.2$\\
69.5 & 73&  $-0.1066 \pm 0.011$ &  $3.6 \%$  &  $-0.119 \pm  0.015 $ & $ 8.5 \pm 0.2$\\
 \hline
\end{tabular}\label{ta2} }
   \end{table}

  We obtain the sizes of  $\rho(s,t=0)$ close to zero and prevalently  positive.
 The contribution of the spin-flip amplitude
    is  sizeable  and impacts  the extracted values of $\rho(s,t=0)$. Most remarkable is
    that the obtained size of the constant of the spin-flip amplitude
    coincides for practically all examined energies.
  These constants are pretty large and have sufficiently small errors.
  This  shows that a careful analysis of the size of $\rho$ requires
 to take into account the contribution of the spin-flip amplitude
  to the differential cross section at small angles, at least at not too high energies.

    The corresponding analysis was made for the proton-proton
    elastic scattering (see Table 3). Again the size of the hadron spin-flip amplitude
    is determined  sufficiently well.
    The changes of the size of $\rho(s,t=0)$ are $5\% \div 10\% $.
  However, as these experimental data have larger errors then in the previous data of the
  proton-antiproton scattering, the determination of the spin-flip amplitude is worst and
  the impact its contribution on the determined size of the $\rho(s,t=0)$ is small.
\section{Additional effects (oscillation)}

            Previously we have  examined various standard  corrections to the exponential behavior
  of the elastic scattering amplitude.
  However, some additional corrections connected with non-standard contributions
  to the main exponential amplitude should be taken into account,
  such as bumps  in the momentum  transfer \cite{Barshay}, or
  small-$t$  additional oscillations in the scattering amplitude \cite{Zarev,Sel-osc}.
  Such oscillations were noted at high energy in the analysis of the UA4/2 Collaboration
  \cite{nic-Sel}.     Recently many other experimental data have been
  analyzed for the presence of  oscillations  \cite{Rev-LHC,Diff10}.
  The analysis points to the possible existence of oscillations in the
 low energy experimental data of the proton-antiproton scattering
  at $3.8 \leq p_L \leq 6.2$ GeV \cite{Diff10}.

  Let us try to simulate  additional oscillations of the elastic scattering amplitude in the form
      \ba
       F^{h}_{osc}(t)= h_{osc} \ sin[\pi \ (\phi_{0}+ q) / q_{0}] \ G_{d}^2(t).
\ea
where  $G_{d}(t)$ is the standard dipole form factor,  $\phi_{0}$ is
the starting phase, and  $q_{0}$ is momentum transfer determining
the period of the oscillation function (order $\sim 0.015 \ $GeV ).
  The largest  contribution  comes from the interference of this
amplitude with the electromagnetic part of the scattering amplitude.
 The inclusion  of such  an additional amplitude essentially decreases  $\chi^2$
 in the analysis of the proton-antiproton   elastic scattering (see $R_{\chi^{2}}$ in  Table 4,
  where $R_{\chi^{2}}$  is defined in eq.(\ref{Rchi}).
    The  decreasing that occurs for proton-proton scattering (see, Table 5)
  less pronounced, but still visible.
  Practically in  all cases  the sizes of the spin-flip amplitude
  and the additional oscillation part are determined well.
    The new values of $\rho(s, t=0)$ for the proton-proton
    scattering lie slightly lower than the experimental data.
    For the proton-antiproton the new sizes of  $\rho(s,
    t=0)$ are slightly above the  experimental data (Fig.2).

\begin{table}
\tbl{ 4   Proton-antiproton elastic scattering (the spin dependence + oscillations)} \\
{\begin{tabular}{@{}|c|c|c|c|c|c|c|c@{}} \\  \hline  
 $p_{L(\frac{GeV}{c})}$ &  N & $\rho_{exp.}$  &$R_{\chi^2}$
 & $\rho_{model}$ & $h_{sf}$, GeV & $h_{osc.} 10^2$, GeV  \\
              &               &               &          &    &  & \\ \hline
3.702 & 134 &$+0.018 $ &  $12.7 \%$  &  $+0.044 \pm 0.01 $ & $34.7 \pm 6.5 $ & $ 9.1 \pm 3.7$\\
4.066 &  34 &$-0.015 $ &  $28.1 \%$  &  $+0.055  \pm 0.02 $ & $44.6 \pm 6.5 $  & $ 9.1 \pm 3.7$\\
5.603 & 215 &$-0.047 $ &  $19.5 \%$  &  $-0.007 \pm 0.005  $ & $ 12.7 \pm 0.6$& $ 11.2 \pm 2.3$\\
5.724 & 115 &$-0.051 $ &  $20.1 \%$  &  $+0.007  \pm 0.005$ & $ 23.1 \pm 0.9$  & $ 13.9 \pm 1.4$\\
5.941 & 140 &$-0.063 $ &  $20.4 \%$  &  $-0.014 \pm 0.002$ & $24.7 \pm 0.6$   & $ 14.5 \pm 0.4$\\
 \hline
\end{tabular}\label{ta3} }
   \end{table}


\begin{table} 
\tbl{ 5   Proton-proton elastic scattering (the spin dependence + oscillations)} \\
{\begin{tabular}{@{}|c|c|c|c|c|c|c|c@{}}  \\  \hline 
 $p_{L(\frac{GeV}{c})}$ &  N & $\rho_{exp.}$  &$R_{\chi^2}$
 & $\rho_{model}$ & $h_{sf}$, GeV & $h_{osc.}10^2$, GeV  \\
              &               &               &          &    &  & \\ \hline
9.4 & 34 & $-0.372 $   &  $1.2 \%$  &  $-0.41 \pm 0.028$ & $ 35.8 \pm 13.7$
& $ 10.8 \pm 4.7$         \\
18.9 & 67 & $-0.266 $  &  $5.8 \%$  &  $-0.28 \pm 0.0117$ & $ 7.8 \pm 1.8$
& $ 7.1 \pm 2.9$\\
38. & 65 & $-0.161 $  &  $2.3 \%$  &  $-0.166 \pm 0.007$ & $ 6.9 \pm 1.7$
& $ 738 \pm 0.5$\\
39.4 & 47&  $-0.195 $ &  $8.7 \%$  &  $-0.19 \pm 0.016$ & $ 12.1 \pm 3.8$
 & $ 7.38 \pm 2.9$\\
40.6 & 65&  $-0.159 $ &  $8.2 \%$  &  $-0.157 \pm 0.011$ & $ 11.5 \pm 4.1$
 & $ 5.9 \pm 2.1$\\
69.5 & 73&  $-0.1066 $ &  $19.3 \%$  &  $-0.176 \pm 0.002$ & $ 8.2 \pm 0.1$
& $ 31.4 \pm 2.4$\\
 \hline
\end{tabular}\label{t6} }
   \end{table}


\label{sec:figures}
\begin{figure}
\begin{center}
\includegraphics[width=0.65\textwidth] {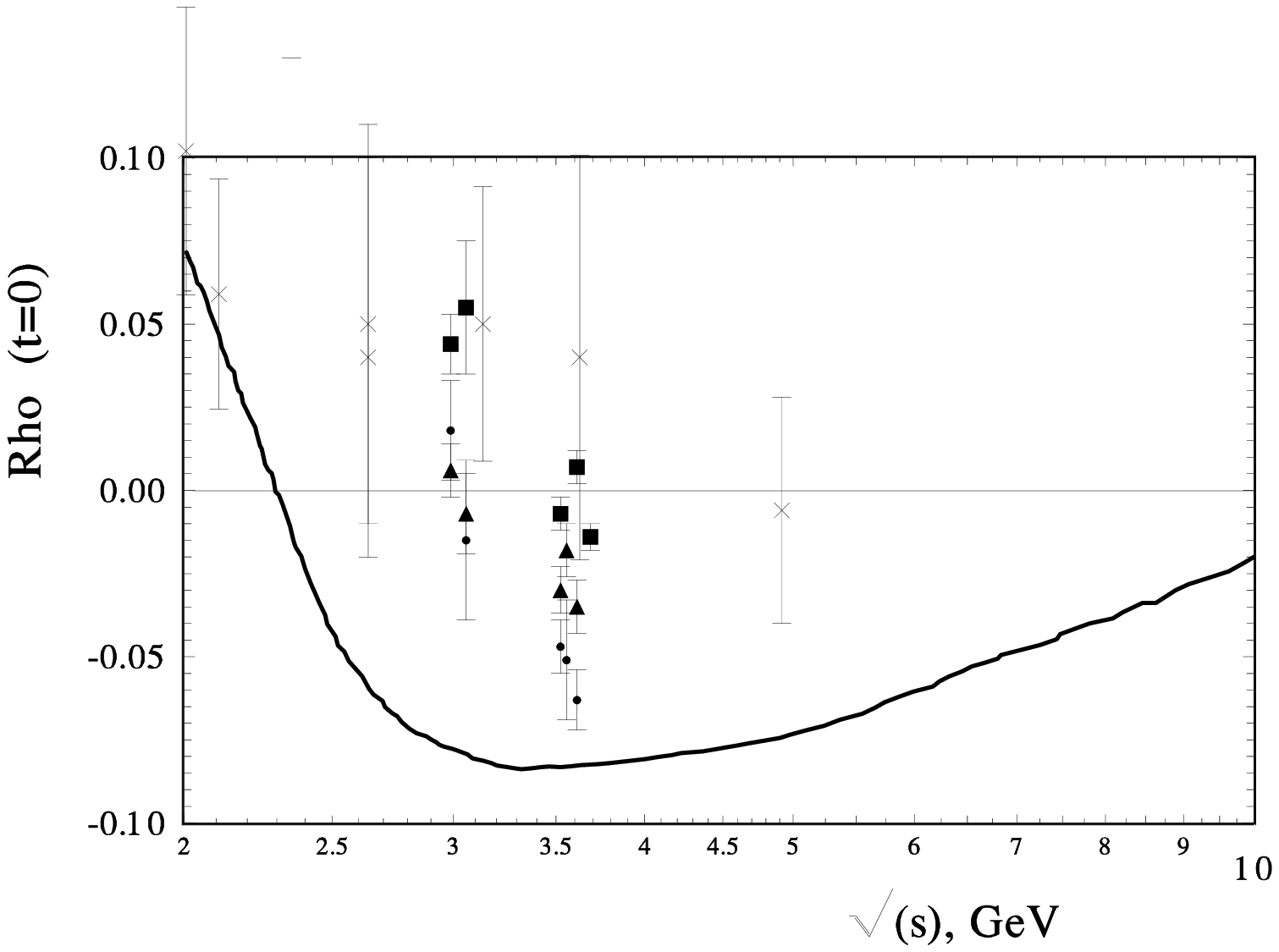}
\end{center}
 \caption{$\rho(s,t=0)$ of the $p\bar{p}$ scattering at low }
 energies (circles and points, as in Fig.1(b), squares - our extraction, see Table 4.
\end{figure}

    \section{Conclusion}

  Future new data from the LHC experiments will hopefully give the possibility to
  carry out new analysis of the  dispersion relation leading to new
  effects such as , for example, a fundamental length of an order of TeV.
   However, for such analysis one also needs the knowledge of the low energy data with
   high accuracy.
 As we have seen, the existing forward-scattering data at $p_L=4 \ \div \ 60  \ $(GeV)
 of the size  $\rho(s,t)$ contradict each other
 and the $\rho(s,t)$ -  data of the proton-antiproton scattering at $p_L=3.7 \ \div \ 6.2 \ $(GeV)
 contradict the dispersion relations prediction in form \cite{Kroll}.
    The present analysis, which includes the contributions
  of the different correction terms in most part support the data of $\rho(s,t=0)$ obtained in \cite{PL-Tro} and show a larger difference with predictions of \cite{Kroll}
   and coincide with some phenomenological analysis based on the dispersion relations carried out in \cite{Lang,Martynov}. However, the last works include in his phenomenological analysis with many free parameters
   for low energy region the data of  the proton-antiproton scattering at $p_L=3.7 \ \div \ 6.2 \ $(GeV).

   It is needed to note that taking into account the contribution of the spin-flip amplitude
   improves the description of the differential cross sections and decreases  $\chi^2$.
    Likely, however,  an additional analysis is needed to include
     additional corrections connected with the possible oscillation in the scattering amplitude and
    with the $t$-dependence of the spin-flip scattering amplitude.
     We hope that
    the forward experiments at NICA can give valuable information for the improvement of our theoretical understanding of the strong hadrons interaction.
This is especially true for the experiment at NICA with polarized beams.

\vspace{0.5cm}
{\bf Acknowledgement}:  {\small \hspace{0.3cm} The author  would  like  to thank
 J.~-R.~Cudell and E. Predazzi for helpful discussions,
 gratefully acknowledges  the financial support
  from BELSPO and would like to  thank the  University of Li\`{e}ge
  where part of this work was done.
   }


\begin{thebibliography}{}

\bibitem{Kroll} P. Kroll, W. Schweiger, Nucl.Phys. {\bf A503} (1089) 865.
\bibitem{Preazzi02} V. Barone, E. Predazzi, "{ \it High Energy Particle Diffraction} ", NY (2002).
\bibitem{mart} A.Martin, F. Cheung, {\it Analytic properties and bounds of the
  scattering amplitude}, (Cordon and Breach, New York, 1970).
\bibitem{roy} S.M.Roy, Phys.Rep. {\bf C 5} (1972) 125 .

\bibitem{Rev-LHC} R. Fiore, L. Jenkovszky, R. Orava, E. Predazzi, A. Prokudin, O. Selyugin,
    Mod.Phys., A24: (2009) 2551.





%

\bibitem{Khuri1} N.N. Khuri, Proceedings of Les Rencontre de Physique de la Villee d'Aoste:
  Results and perspectives in Particle Physics (M. Greeco ed.), p.701 Gif-nur-Yvette, France (1994).
\bibitem{Khuri2} C. Bourrely, N.N. Khuri, A. Martin, J. Soffer, T.T. Wu, Proceedings EDS 2005, Blois, France (2005).







\bibitem{wak} M.Sawamoto, S.Wakaizumi,
      Proc Theor.Phis. {\bf 62} (1979) p.1293.

\bibitem{SCP-EPJ08}
    O.V. Selyugin, J.-R. Cudell, E. Predazzi,
 Eur.Phys.J.ST {\bf 162}, (2008)  37-42.



\bibitem{SelyuginPL94} O.~Selyugin,
 Phys. Lett.{} {\bf B333},~245~(1994)
\bibitem{SelyuginYF92}
O.~Selyugin, Sov. J. Nucl. Phys.{} {\bf 55},~466~(1992)\relax



\bibitem{COMPETE1a}J.~R.~Cudell {\it et al.} [COMPETE Collaboration],
Phys.\ Rev.\ D {\bf 65} (2002) 074024.
\bibitem{COMPETE1b} J.~R.~Cudell {\it et al.} [COMPETE Collaboration],
Phys.\ Rev.\ Lett.\ {\bf 89} (2002) 201801.
\bibitem{COMPETE2} J.~-R.~Cudell, V.~Ezhela, K.~Kang, S.~Lugovsky
and N.~Tkachenko, Phys.\ Rev.\ D {\bf 61} (2000) 034019
[Erratum-ibid.\ D {\bf 63} (2001) 059901]
[arXiv:hep-ph/9908218].


\bibitem{Kuznetzov1} D. Gross, et al., Phys.Rev.Lett. {\bf 41},  (1978) 217.
\bibitem{Kuznetzov2} A.A. Kuznetzov et al., preprint JINR P1-80-376, Dubna (1980).
 \bibitem{Fajardo} L.A. Fajardo et al., Phys.Rev., {\bf D24} (1981) 46.


\bibitem{PL-Tro} T.A. Armstrong et al., Phys.Lett. B {\bf 385} (1996) 479.

\bibitem{bgl}
     N. H. Buttimore, E. Gotsman, E. Leader,  Phys. Rev. D {\bf 35},
         (1987) 407.




\bibitem{akhi}
A.I. Akhiezer, I.Ya. Pomeranchuk, {\it J. Phys.} {\bf 9} (1945) 471.

\bibitem{bethe} H. Bethe,  Ann. Phys. {\bf 3}, (1958) 190;
\bibitem{solov}
  L.D. Soloviev, Zh.Eksp.Teor.Fiz., {\bf 49},(1965) 292.
 \bibitem{rix}
 J. Rix, R.M. Thaler, Phys. Rev., {\bf 152}, (1966)1357.


\bibitem{loch}
 M.P. Locher, Nucl.Phys., {\bf B 2}, (1967) 525.
%
\bibitem{wy}
  G. B. West, D. R. Yennie,  Phis. Rev. {\bf 172}, (1968) 1414.
\bibitem{can}
      R. Cahn, Zeitschr. fur Phys.  {\bf C 15}, (1982) 253.

 \bibitem{selmp1} O.V. Selyugin, Mod. Phys. Lett. A{\bf 11} (1996) 2317.
\bibitem{selmp2} O.V. Selyugin, Mod. Phys. Lett. A{\bf 12},  (1997) 1379.
\bibitem{PRD-Sum}
   O.\,V.~Selyugin,
  Phys.\ Rev.\  D {\bf 60} (1999) 074028


 \bibitem{lap} L.I. Lapidus, Particles \& Nuclei {\bf 9}, (1978) 84.




\bibitem{Disser} S. Trokenheim, {\it Fermilab-Thesis-1995-40},(1995).
\bibitem{Disser-data} Durham~HepData~Project, M.R. Whalley,
http://durpdg.dur.ac.uk/hepdata/reac.html.
\bibitem{Land-Bron} K.R. Schubert, In Landolt-B$\ddot{o}$rnstein, New Series, v. 1/9a, (1979).

\bibitem{Dubna-11}  J.-R. Cudell, E. Predazzi, and O.V. Selyugin,
Proceedings of the Int.Conf. "High energy spin physics",  ed. A.V. Efremov,
        Dubna, (2011).

\bibitem{Sel-spin}
O.V.~Selyugin,
Proceedings of the Int.Conf. "New Trends in High Energy Physics",
 ed. P.N.Bogolyubov, L.Jenkovszkky,Yalta(Crimea), September 22-29, 2001, Kiev (2001), p. 237.



\bibitem{PS-EPA02}  E. Predazzi, O.V.~Selyugin,
{\it Eur.Phys.J.}  {\bf A 13 }, (2002) 471--475.


\bibitem{CPS-EPA04}  J.-R. Cudell, E. Predazzi, and O.V. Selyugin,
          Eur. Phys. J. {\bf A 21} (2004) 479--486.




\bibitem{sum-L} N.H. Buttimore et al.,   Phys.Rev. D {\bf 59} (1999) 114010.
  \bibitem{soff}
  C. Bourrely, J. Soffer, hep-ph/9611234.



\bibitem{M-Pred} A.F. Martini, and E. Predazzi, Phys. Rev. D  {\bf 66} (2002) 034029.

\bibitem{slope-MPL99} O.V. Selyugin,
          {\it Mod.Phys.Lett, {\bf A 14},  (1999)} 223.



\bibitem{mog2a}
B.Z. Kopeliovich and B.G. Zakharov, Phys.lett. {\bf B226}  (1989) 156.
\bibitem{mog2b} M. Anselmino and S. Forte, Phys. Rev. Lett. {\bf 71}, (1993) 223.

\bibitem{mog2c}  A.E. Dorokhov, N.I. Kochelev and Yu.A. Zubov,
Int. Jour. Mod. Phys. {\bf A8},  (1993) 603.
\bibitem{mog2d} N. Akchurin, S.V. Goloskokov, O.V. Selyugin, Int. J. Mod. Phys. A {\bf 14}
   (1999) 252.

\bibitem{CPS-PN04}J.R. Cudell, E. Predazzi, O.V. Selyugin,
          Particles\&Nuclei, {\bf 36(7)} (2004) 132.
 \bibitem{yaf-wak}
 O.V.Selyugin,
  Phys. of Atomic Nuclei {\bf 62} (1999) 333.

\bibitem{predaz1} E. Predazzi, G. Soliani, Nuovo Cim. {\bf A 2} (1967) 427.
 \bibitem{predaz2}   K. Hinotani, H.A. Neal, E. Predazzi and G. Walters,
   Nuovo Cim., {\bf A 52} (1979) 363.



\bibitem{Runco} S.B. Nurushev and  V.A. Okorokov, Proceedings XII Workshop on high energy spin physics,
   DSPIN-2007, Sept. 3-7, Dubna (2007), Dubna, 2008, p. 117.
\bibitem{Barshay} S. Barshay, P. Heiliger, {\it Z. Phys.} {\bf C 64} 675 (1994).

 \bibitem{Zarev}   N.~I.~Starkov, V.~A.~Tsarev,
  Pisma Zh.\ Eksp.\ Teor.\ Fiz.\  {\bf 23}, (1976) 403.

          \bibitem{Sel-osc}  O.V. Selyugin, in {\it Hadrons-95}, Proc. of the XIth
         Workshop  on Soft Physics, edited by L. Jenkovszky, Kiev,  (1995), p.65.

   \bibitem{nic-Sel}P. Gauron, B. Nicolescu, O.V. Selyugin,
    Phys.Lett.  {\bf 397} 305  (1997).

  \bibitem{Diff10} O.V. Selyugin, J.-R. Cudell, in the International Workshop
          on Diffraction in High-Energy Physics, Otranto (Lecce, Italy), September 10-15,
          2010; AIP Conf. Proc. 1350:115-118, 2011; ArXiv:hep-ph/1011.4177.
 \bibitem{Lang} J.R. Cudell, E. Martynov, O.Selyugin, A. Lengyel,
              Phys.Lett. B {\bf 587} 78 (2004).
                        \bibitem{Martynov} J.R. Cudell, E. Martynov, O.Selyugin,
              EPJ C{\bf 33} s 533 (2004).

\end{thebibliography}
\end{document}